# Dynamic coexistence of various configurations: clusters vs nuclei

Vladimir Z. Kresin [(1)*] and Jacques Friedel [(2)]

[(1)] Lawrence Berkeley Laboratory, University of California

at Berkeley, CA 94720, USA

[(2)] Laboratoire de Physique des Solides, University

Paris XI, Batiment 510, 91405 Orsay Cedex, France

## Abstract.

The presence of energy shells in metallic clusters and atomic nuclei leads to a peculiar relation between the number of particles N and the structure, and this leads to a strong correlation between the energy spectrum and N. An analysis of experimental data leads to the conclusion that, in addition to the static Jahn-Teller effect, the *dynamic* effect leading to the quantum coexistence of different configurations (quantum oscillations) plays an important role. Such suggested coexistence is an essential feature of clusters as well as nuclei, both finite Fermi systems.

## 1. Introduction.

The electronic states of metallic nanoclusters form energy shells [1] similar to those in atoms or nuclei , see ,e.g., reviews [2-4] . The cluster configuration depends on the degree of shell filling . As a result, the cluster shape, and, therefore, the electronic energy spectrum depend on the number of delocalized electrons N. A similar dependence on the number of nucleons is observed for atomic nuclei (see, *e.g.,* [ 5 ]). The present paper is concerned with this remarkable dependence.

Clusters with filled shells ("magic" clusters) possess approximately spherical symmetry. The quantum states of "magic" clusters are classified by values of the orbital momenta. As for the clusters with incomplete shells , their orbital degeneracy leads to an instability of the spherical shape. Therefore, adding or removing electrons from the "magic" cluster leads to change in the cluster geometry, and this change in turn affects the electronic energy spectrum.

An analysis of the evolution of the cluster shape is not a trivial task. The generally accepted view (see, *e.g.,* [2]) is that , in first approximation, the cluster with incomplete shell has an spheroidal structure. Such a deviation from the spherical shape is caused by the electron-vibrational interaction, that is, by the static Jahn-Teller (JT) effect which removes the orbital degeneracy. The initial deformation caused by adding one or more electrons leads to prolate configuration ,but after filling more than half-shell the shape changes to oblate. This point of view is supported by calculations for small metallic clusters performed with use of the jellium model [6]. The conclusion is based on energy consideration and the configuration with smaller energy is accepted as the real one. However, such a picture is a bit oversimplified. First, the pure prolate as well as oblate electronic states are classified by projection of the orbital momentum "m" and are doubly degenerate; this should lead to an additional deformation, and as a whole there should be a triaxial distortion. However, usually the real picture is different and, as will be discussed below, is dynamic. The question about such an evolution of metallic clusters is directly related to an analysis of their optical absorption spectra. There is a direct correlation between these spectra and the cluster configuration. As is known, these spectra are peaked at the cluster plasmon frequencies(see, *e.g.,*[2]).

Correspondingly, for the spherical "magic" clusters, one peak is observed. As for the clusters with incomplete energy shells, one could expect triaxial JT static deformation and, as a result, three-peak structure of the spectra. However, experimentally this is not the usual case for clusters (see below, Sec. 3 ), and this should be explained. In addition, the dependence of the amplitudes of the peaks on N looks rather peculiar.

It will be shown, that, in competition with the static JT effect, one should take also into account the dynamic phenomenon, namely the quantum transitions (oscillations) between the quasi-resonance states , and it leads to the coexistence of different structures. One should stress that such a coexistence is different from the picture of a mixture of different isomers, which are isolated clusters in stationary states. The last situation is a limiting case of the concept discussed here (see below,Sec.2).

On a whole, the picture looks analogous to that in nuclear physics. Indeed, the studies of various nuclei, including recent experiments

( see, *e.g.,* [ 7-12 ]) have demonstrated the coexistence of various configurations. We would like to point out that the dynamic phenomenon described below is manifested in both systems (clusters and nuclei).

The structure of the paper is as follows. Sec.2 describes a general picture of the coexistence of various configurations in nanoclusters and an analogy with the situation in nuclear physics. The description of this concept is the key ingredient of this paper. The problem of dynamic coexistence vs. static Jahn-Teller effects is also discussed . Sec.3 contains an analysis of the experimental data and general discussion.

## 2. Configurations and their coexistence.

The shell structure of clusters in many aspects is similar to that in atomic nuclei. The shape of a nucleus can be described by the following expression (see, *e.g.,* [ 5 ]):

$$R(\theta,\varphi) = R_0 \{1+\beta (5/16\pi)^{1/2} [\cos\gamma (3\cos^2\theta-1)+ \\ 3^{1/2}\sin\gamma \sin^2\theta \cos2\varphi]\} \quad (1)$$

Here $R(\theta,\varphi)$ is the length of the vector pointing from the center to the surface, $R_0$ is the radius of the spherical nucleus, $\theta$ and $\varphi$ are the spherical coordinates for the fixed coordinate system; Eq.(1) describes the most interesting case of quadrupole deformations. The quantities $\gamma$ and $\beta$ are the so-called Hill-Wheeler coordinates [13]. These coordinates describe the scale of deformation ($\beta$) and the degree of mixture of the prolate and oblate configurations ($\gamma$). The value $\gamma=0$ corresponds to the prolate configuration ( the axis z has been chosen to be the symmetrical axis), see eq.(1), and $\gamma=\pi/3$ describes the oblate configuration (with x as the axis of symmetry), see [5,11] and fig.1. Intermediate values of $\gamma$ correspond to triaxial deformation.

Eq.(1) can be also used to describe the geometry of metallic clusters. It is convenient to write it in the form:

$$R = R_0\{1 + \beta(5/16\pi)^{1/2}[(\cos\gamma - 3^{-1/2}\sin\gamma)f_{pr.} + (2/3^{-1/2})\sin\gamma f_{obl.}]\}$$

(2)

where

$$f_{pr.} = (3\cos^2\theta - 1)$$

and

$$f_{obl.} = 0.5(3\cos^2\theta - 1) + 1.5\sin^2\theta \cos 2\varphi.$$

Indeed, one can see that the values $\gamma=0$, $\pi/3$ correspond to the prolate and oblate configurations, correspondingly. Note that at $\gamma=\pi/6$ both configurations make equal contributions.

Of course, there are specific features of clusters different from those for nuclei. First, the clusters contain just one type of fermions (electrons). The cluster stability is provided by Coulomb forces. Moreover, the presence of heavy ions allows us to employ the adiabatic approximation; as a result, one can introduce the classification of energy levels and the interaction ( non-adiabaticity ) . One should keep in mind that the analogy between clusters and nuclei should be pursued with care, because a microscopic analysis is rather different.

Therefore, the shape evolution of clusters can be described in a way which combines the analysis similar to that in nuclear physics along with picture, based on the adiabatic approximation. The most important parameter is $\gamma$ (see eqs.(1),(2)) which describes contributions of various configurations.

According to the picture based on the static JT effect, the progressive filling of the energy shells produces successive prolate, triaxial, and oblate configurations. However, based on microscopic theory with use of the so-called diabatic approximation, one can introduce an alternative scenario [14]. The shape of the cluster with incomplete shell can be viewed as a quantum superposition (coexistence) of prolate and oblate configurations ( fig.1). For slightly occupied shells the prolate configuration is dominant, whereas the oblate configuration is dominant for the opposite case of a nearly full shell. However, the picture of dynamic superposition introduced here is important near the half-shell filling and in the intermediate region.

More precisely, let us focus on the clusters with an upper shell being less than half-filled, that is $N<N_m/2$, where N is the number of the delocalized electrons in the incomplete shell, and $N_m$ is the total number of states in this shell. The wave function can be written in the form:

$$\Psi(\vec{r},\vec{R},t) = a(t)\varphi_a(\vec{r},\vec{R}) + b(t)\varphi_b(\vec{r},\vec{R}) \qquad (3)$$

Here a,b correspond to prolate and oblate configurations, $\{\vec{r},\vec{R}\}$ are the electronic and ionic coordinates. Assume that a(0)=1, b(0)=0, that is, at t=0 the cluster is in the prolate configuration. It is convenient to employ the diabatic representation described in [15,16]. In usual adiabatic picture we are dealing with a single potential energy surface (PES; sometimes it is called the adiabatic potential, or the energy term) with two minima. The transition to the diabatic representation means that one can introduce two crossing PES's corresponding to the prolate and oblate structures. It is essential that, contrary to the usual adiabatic scenario, the operator $\hat{H}_{\vec{r}} = \hat{T}_{\vec{r}} + V(\vec{r},\vec{R})$ is not diagonal in the diabatic representation ($\hat{T}_{\vec{r}}$ is the operator of kinetic energy of electrons, and $V(\vec{r},\vec{R})$ is the total potential energy). The wave function $\varphi_{a(b)}(\vec{r},\vec{R})$ has a form:

$\varphi_{a(b)}(\vec{r},\vec{R}) = \tilde{\psi}_{a(b)}(\vec{r},\vec{R})\tilde{\phi}_{a(b)}(\vec{r})$, where $\tilde{\psi}_{a(b)}(\vec{r},\vec{R})$ and $\tilde{\phi}_{a(b)}(\vec{R})$ are the electronic and vibrational wave functions in the diabatic representation. Using time-dependent Schrödinger equation (see, *e.g.,* [17] ) and eq.(3), we obtain :

$$|b(t)|^2 = (\varepsilon_{ab}^2/2)\left[(\Delta\varepsilon/2)^2 + \varepsilon_{ab}^2\right]^{-1}\left\{1 - \cos 2\Omega t\right\}$$

$$\Omega = \left[(\Delta\varepsilon/2)^2 + \varepsilon_{ab}^2\right]^{1/2} \tag{4}$$

Here $\Delta\varepsilon = \varepsilon_b - \varepsilon_a$

$$\varepsilon_i = \int d\vec{R}\,\tilde{\phi}_i \left[\hat{T}_{\vec{R}} + \hat{H}_{\vec{r};ii}(\vec{R})\right]\tilde{\phi}_i$$

$$\varepsilon_{ab} = \int d\vec{R}\,\tilde{\phi}_a \left[\hat{T}_{\vec{R}} + \hat{H}_{\vec{r};ab}(\vec{R})\right]\tilde{\phi}_b$$

$$\hat{H}_{\vec{r};ab}(\vec{R}) = \int \tilde{\psi}_b^* \hat{H}_{\vec{r}} \tilde{\psi}_a d\vec{r}$$

One can see from eq.(4) that the cluster oscillates between the prolate and oblate configurations. The frequency of oscillations is determined by the transition parameter $\varepsilon_{ab}$ and by the energy difference $\Delta\varepsilon$. The transition parameter describes the probability of tunneling through the barrier separating two configuration. Variation of this parameter directly correlate with the experimental measurements of the absorption spectra (see below, Sec.3). The oscillations (4) are similar to those for the benzene molecule between the Kekule configurations ( see, *e.g.,* [18]), described in the Heitler-London approximation.

Note also that the transition parameter $\varepsilon_{ab}$ can be written in the form :

$$\varepsilon_{ab} = L_0 F \tag{5}$$

where

$$L_0 = \int d\vec{r}\, \tilde{\psi}_b^*(\vec{r},\vec{R}) \hat{H}_{\vec{r}} \tilde{\psi}_a(\vec{r},\vec{R})_{|\vec{R}=\vec{R}_0} \tag{5'}$$

is the electronic parameter, and

$$F = \int d\vec{R}\, \tilde{\phi}_b(\vec{R}) \tilde{\phi}_a(\vec{R}) \tag{5''}$$

is the Franck-Condon factor; its value depends on the overlap of the vibrational wave functions for the prolate and oblate configurations, $\vec{R}_0$ corresponds to the crossing region.

Therefore, adding of electrons to the "magic" cluster distorts its spherical shape and this produces the oscillations described by eq.(4). We are dealing with the dynamic process similar to that which in molecular spectroscopy is called the dynamic Jahn-Teller effect , or the tunneling splitting ( see, *e.g.,* [19] , [20] ). Note that the usual static JT effect is caused by the electron-lattice interaction. This interaction, described by the Hamiltonian which is linear in the ionic displacement, leads to removal of the degeneracy. In the dynamic scenario the coexistence of different configurations is described by a different Hamiltonian, namely by the term $\varepsilon_{ab}$ (in the diabatic representation), see eq.(4). This term plays a role, similar to that of a tunnelling Hamiltonian ( see, *e.g.,* [21,22]).

The dynamic coexistence introduced here is different from the picture of isomers. The last situation is a limiting case of the proposed scenario and corresponds to a negligibly small value of the transition parameter $\varepsilon_{ab}$.

One can see directly from eq.(4) that the contribution of the oblate state determined by the value of the b(t) depends on the transition parameter $\varepsilon_{ab}$ and the energy difference $\Delta\varepsilon$. If the transition probability is very small (e.g., $\varepsilon_{ab} \ll \Delta\varepsilon$), then the cluster essentially keeps its initial prolate state (recall that we focus on the case when $N<N_m/2$). Such a situation corresponds to the static Jahn-Teller effect. The picture is different if $\varepsilon_{ab} \gtrsim \Delta\varepsilon$. It is possible to observe also an intermediate situation. Its existence is essential, because it provides a gradual transition between the regions $\varepsilon_{ab} \ll \Delta\varepsilon$ and $\varepsilon_{ab} \gtrsim \Delta\varepsilon$. As a result, continuing shell filling does not produce a sudden jump between the case of definite prolate configuration and the case of oscillations between two shapes. In actuality, the amplitude of the oblate configuration is simply undergoing a gradual increase upon filling. This type of analysis allows us to explain seemingly paradoxical aspects of experimental data on optical absorption (see below, Sec.3).

Consider again eq.(4) . One can obtain:

$$\overline{b}^2 = 0.5 r^2 [1+r^2]^{-1} \qquad (6)$$

where

$r = \varepsilon_{ab}/(\Delta\varepsilon/2)$

$\overline{b}^2$ describes the average contribution of the oblate structure. Therefore the cluster structure, similar to nuclei, is a superposition of the prolate and oblate configurations . Similarly to the notation in nuclear physics, one can introduce the parameter $\gamma$ (cf. eqs.(1),(2)) , so that

$$\sin \gamma = 0.5\, r^2(1+r^2)^{-1} \qquad (7)$$

Note that if $\Delta\varepsilon \gg \varepsilon_{ab}$, then $\gamma \approx 0$, and the prolate configuration is dominant. Then b(t) is negligibly small; in this case the result coincides with that obtained from the picture of the static JT effect (see above). Near the half-shell filling the configurations have very close values of energy ( see, *e.g.,* [6] ), so that $\Delta\varepsilon<<\varepsilon_{ab}$. Then $\bar{b}^2 \approx \bar{a}^2 = 0.5$, and $\gamma \approx \pi/6$ ( cf. eq. ( 2 )).

One can conclude that the cluster structure is determined by the ratio $\varepsilon_{ab}/\Delta\varepsilon$. Since $\varepsilon_{ab} \propto F$, the value of the Franck-Condon factor is an essential ingredient determining the cluster geometry and, correspondingly, its spectroscopy.

Therefore, the cluster structure represents a superposition of two configurations. A similar picture has been discussed for atomic nuclei ( [7]-[12], see below) in the static approximation, where for nearly half filled shells triaxial deformation has been considered.

### 3. Experimental data. Discussion.

The shape of metallic clusters is manifested in the spectrum of collective electronic modes, surface plasmons. As is known (see, *e.g.,* review [ 23 ]), the photoabsorption spectrum is peaked at the plasmon frequency. As was noted above, for spherical clusters one can observe a single peak at $\omega=\omega_{pl}$. However, for clusters with incomplete shells the absorption spectrum has a different structure, and this is caused by a change in the cluster shape.

Let us discuss the experimental data on the clusters absorption. If the cluster contains an incomplete energy shell, it undergoes the Jahn-Teller distortion removing the orbital degeneracy. One can expect that such a cluster should have a static triaxial shape. In this case one might expect the adsorption spectrum to have three plasmon peaks. However, usually this is not the case. The detailed measurements [24] of spectra of sodium clusters (14-48 atoms) demonstrated that a majority of them display two peaks; a few of them are characterized by only a single peak.

In addition, there is another feature which must be also explained. Consider a cluster with a prolate configuration; its spectrum, indeed, is characterized by the presence of two peaks, and the frequencies $\omega_z$ and $\omega_x=\omega_y$ are different (the axis z has been chosen to be along the symmetry axis) . Since the frequency along the x direction is doubly degenerate, one should expect the ratios of the oscillator strengths to be equal to 2:1. Therefore, the ratio of the intensities of two peaks in the absorption spectrum should be 2:1. However, this is observed only for clusters with slightly occupied shells. As N evolves towards half-filling, the peak ratio changes (fig.2) and then becomes almost 1:1.

Original rationalization of the deviations from the 2:1 ratio was based on qualitatively referring to the static picture of a triaxial deformation with two close peaks, see, e.g., [25]. However, it cannot easily explain the general trend toward the ratio 1:1 appearing with increased shell filling.

The increase in filling leads to the ratio 1:1.One can show that these seemingly paradoxical experimental features can be explained as manifestations of the dynamic effect ( resonance between quasi-resonance states), described above.

Indeed, let us discuss the data [ 24 ] in more detail. The ion $Na_{21}^+$ containing N=20 delocalized electrons   ( as well as $Na_{41}^+$ ; N=$N_{mag.}$=40)  is a "magic" cluster.  Because of  spherical shape , its plasmon frequency has a single value. This correlates with a single peak in the absorption spectrum. Increase in N leads to an appearance of two peaks structure observed in [24]. According to our approach, this phenomenon can be explained by the coexistence of two configurations (prolate and oblate).

Initially, the prolate structure dominates and the ratio of amplitudes is close to 2:1; the picture is similar to that for the static Jahn-Teller effect. Subsequent increase in N leads to an increase in the contribution of the oblate configuration and to a decrease in the ratio. According to [24], the cluster $N_{30}^+$ has two almost equivalent peaks. This is in a good agreement with our theory, since this cluster contains almost half-filled shell. The further increase in N leads to the oblate configuration becoming dominant, and it is reflected in the relative values of the amplitudes.

There is one exceptional case, namely, the cluster $Na_{31}^+$. This cluster is not "magic"; but, nevertheless, the structure observed in [24] contains only a single peak ; a similar picture for this cluster was observed also in [25]. One should note also that the width of the peak is larger than that for "magic" cluster. This observation can be explained in the following way. This cluster corresponds to the prolate→oblate transition. Because of it, the deformation is rather small, so that both configurations, prolate and oblate, are close to spherical. As a result, the values of plasmon peaks frequencies are relatively close. In addition, because $\Delta\varepsilon \approx 0$, and because of small value of $\varepsilon_{ab}$ (see fig.1 in [14]) the value of the frequency $\Omega$ is also small (see eq.(4)). Because of it, this case is dominated by the static JT effect, that is we are dealing with the triaxial deformation and, correspondingly, the plasmon spectrum has three peaks; however, because of small scale of deformation, the difference between the values of the plasmon frequencies for these peaks $\Delta\omega_{pl.}$ is small. This smallness leads to the absorption spectrum displaying a single peak structure with its width which is larger than $\Delta\omega_{pl}$ ( $\Delta\varepsilon \gtrsim \Delta\omega_{pl}$ ). However, this width is larger than that for the "magic" clusters (e.g., for $Na_{21}^+$ and $Na_{41}^+$)  This conclusion corresponds to the observation for $Na_{31}^+$ [24], [25]. On the whole, this special case deserves a more detailed study.

As was noted above, the theoretical and experimental study of shape coexistence in nuclei is undergoing a very intensive development (see, *e.g.,* [7-12,26,27]). This problem appears to be exceptionally important for the nuclei which are far from the stability line, for example, for neutron rich nuclei. Recent measurements of electromagnetic moments [12] provide unambiguous evidence of coexistence of different shapes for $^{43}S_{27}$ nuclei. The nuclei as well as nano-clusters are characterized by the presence of energy shells ( see, *e.g.,* [2], [5], [28] )."Magic" nuclei have a spherical shape, and the nuclei with incomplete shells undergo distortion similar to that in clusters. Note also that the presence of shell structure leads to superconducting pairing. This is a well-known concept in nuclear physics. For metallic clusters a similar phenomenon was studied theoretically in our papers [29], [30].

The shape coexistence seems manifested for both systems. One should expect that the shape coexistence in nuclei, similar to clusters ,is manifested in dynamic as well in static effects, and there could be also an intermediate situation.

Note that even for "magic" clusters one can observe a small splitting of the peak (see, *e.g.,* [31,25]). This feature can be explained by an overlap of collecive and single-particles energy manifolds [ 32] . However, the coexistence of spherical and deformed configurations can also contribute to the splitting. In connection with this, it is interesting that such a coexistence was determined for some nuclei (*e.g.,* for $^{13}S_{27}$ [12 ]), see also [27]. This coexistence corresponds to two close potential minima (see [12], fig.4).

For larger clusters (N ≳ 40) the situation becomes more complex. This is due to the fact that the splitting caused by the deformation, might lead to the energy overlap of the states from different shells. The shape coexistence, caused by the interplay of the dynamic scenario and the static JT effect is also manifested in such clusters, but the analysis is more complicated. A more detailed discussion will be presented elsewhere.

Finally, speaking of absorption spectra, let us consider the experimentally measured quantity B defined as

$$B = h^{(1)}/h^{(2)} - 0.5 \qquad (8)$$

Here $h^{(1)}$ and $h^{(2)}$ are the intensities, that is the areas under the lower $\omega_{pl.;z}$ and the higher ($\omega_{pl.;x} = \omega_{pl.;y}$) frequencies. The value of B describes the contribution of the oblate configuration. More specifically, $B \approx \bar{b}^2$ ( see above, eq.(7) ). Correspondingly, $B \approx 0$ for slightly occupied shells, and $B \approx 0.5$ for the clusters with a nearly half filled shell.

With use of eqs.( 7 ),( 8 ), one can write :

$$\sin \gamma \approx B \qquad (9)$$

Therefore, the parameter $\gamma$ describing the mixing of the prolate and oblate configurations, can be expressed through the experimentally measured quantity B, defined by eq.(8).

In summary, the presence of shell structure of spectra for metallic nanoclusters allows us to introduce the concept of dynamic coexistence of various shapes (prolate and oblate), that is the picture of quantum oscillations between the configurations. This possibility should be explored more carefully in nuclei. The puzzling experimental data on optical absorption and its dependence of the shell filling can be explained by this dynamic phenomenon.

The study of coexistence of various shapes is an interesting direction in nanophysics.

The authors are grateful to V.V. Kresin and M. Kaplan for interesting discussions The research of VZK was supported by AFOSR.

. **Figure captions**

Fig.1. Polar plane ($\gamma,\beta$) for the description of various shapes. The $60^0$ wedge desribes all possible shapes as superpositions of the prolate and oblate configurations.

Fig.2. Optical absorption spectra for the "magic' cluster $Na_{21}^+$ and clusters $Na_{28}^+$ and $Na_{29}^+$ (ref.24); $\sigma$ is the cross section per electron.

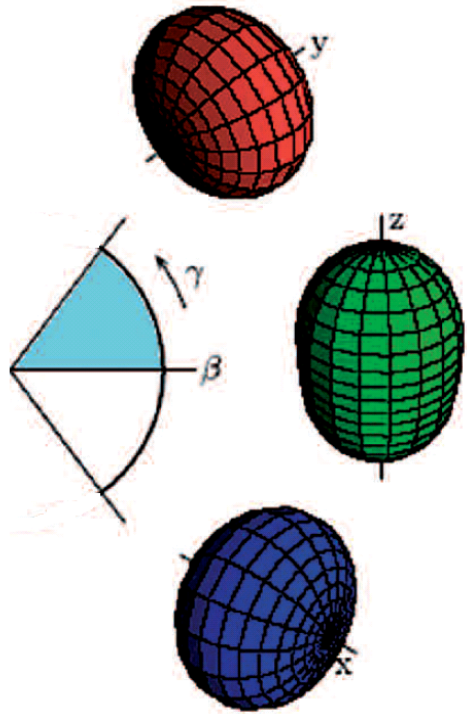

Fig.1

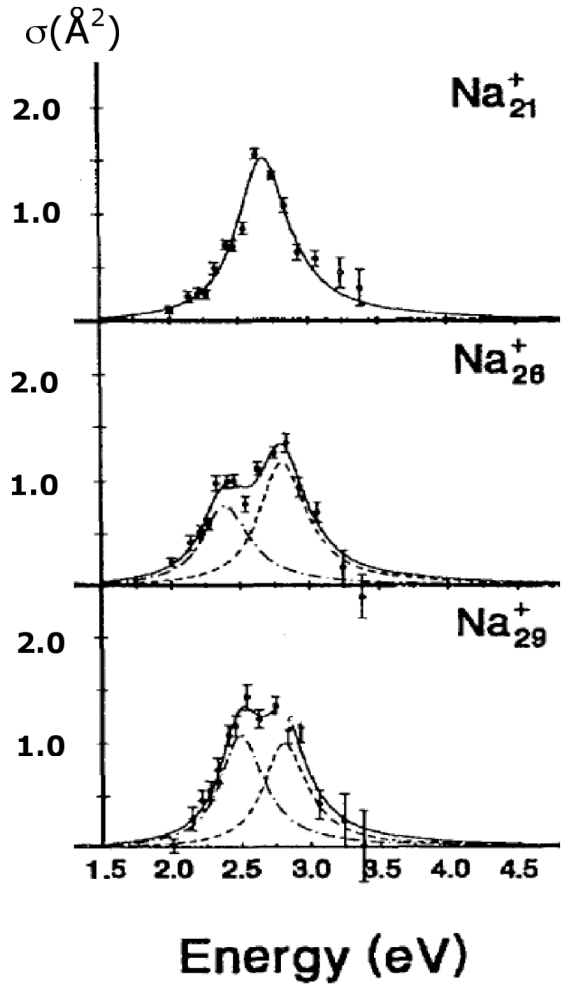

Fig.2